# On the Evaluation of Military Simulations: Towards A Taxonomy of Assessment Criteria

Mario Golling[1,2,3], Robert Koch[1,2,3], Peter Hillmann[3], Volker Eiseler[3], Lars Stiemert[3], Andres Rekker[1,2]

[1] Helmut-Schmidt-Universität
Universität der Bundeswehr Hamburg
Holstenhofweg 83
22043 Hamburg
Email: {mario.golling,
robert.koch,andres.rekker}@hsu-hh.de

[2] General Staff College
Clausewitz-Kaserne
Manteuffelstraße 20,
22587 Hamburg, Germany

[3] Universität der Bundeswehr München
Werner-Heisenberg-Weg 39,
85577 Neubiberg, Germany
Email: {peter.hillmann, volker.eiseler,
lars.stiemert}@unibw.de

*Abstract*—In the area of military simulations, a multitude of different approaches is available. "Close Combat Tactical Trainer", "Joint Tactical Combat Training System", "Battle Force Tactical Training" or "Warfighter's Simulation 2000" are just some examples within the history of the large DoD Development Program in Modelling and Simulation, representing just a small piece of the variety of diverse solutions. Very often, individual simulators are very unique and so it is often difficult to classify military simulations even for experienced users. This circumstance is further boosted due to the fact that in the field of military simulations – unlike in other areas - no general classification for military simulations exists. To address this shortcoming, this publication is dedicated to the idea of providing a first contribution to the development of a commonly accepted taxonomy in the area of military simulations. To this end, the problem field is structured into three main categories (general functional requirements for simulators, special military requirements for simulators and non-functional requirements for simulators). Based upon that, individual categories are provided with appropriate classes. For a better understanding, the taxonomy is also applied to a concrete example (NetLogo Rebellion).

## I. INTRODUCTION

At least since 1824 when Karl von Müffling convinced the Prussian Army of the fundamental advantages of his "war game" ("Kriegsspiel"), war games and military simulations have been used in various forms within military training. Based on the idea that they are well suited to "improve the tactical and strategic skills", such simulations - albeit in modernized forms – are still in use at military academies/military organizations worldwide as well as in so-called think tanks, among others for the purpose of conflict analysis [1]. With their (i) capacity of dynamically representing the physical world and (ii) the ability to apply the lessons learned to real world situations, war games are very interesting and useful tools in order to plan for future courses of action and decision making as well as for the development of problem solving techniques [2]. Related to the issue of conflict simulations, war games provide an effective way to simulate historical, current or future conflicts as well as the consequences of making decisions.

### A. Goal of the Paper

Over the last two centenaries a great variety of different military simulations has arisen, which makes a comparison very difficult (see [3] for a comprehensive treatment, including an extensive review of the history and development of the subject). To enable such comparisons (amongst other reasons) taxonomies are used. Such taxonomies are the hierarchical structuring of a knowledge field into main groups and subcategories [4]. Based on the allocation of the taxonomy, the nature of the different classes should be understood in detail (e.g, see [5], [6], [7], [8] for an excerpt of taxonomies in the field of computer / network security). In contrast to other fields of research where generally accepted taxonomies exist, in the field of military simulations - due to the absence of a generally accepted classification - a comparative analysis (assessment) is currently very difficult. Therefore, this paper is dedicated to the introduction of a taxonomy for military simulations.

### B. Outline of the Paper

For this, Section 2 gives an overview of important terms, which are relevant for the overall understanding. Based on that, Section 3 describes the taxonomy with all its classes. Thereafter, Section 4, contains a small case study where our taxonomy is applied to (here, NetLogo Rebellion). Finally, Section 5 summarizes the key results and gives an outlook on possible future developments.

## II. DEFINITION OF TERMS

Before going deeper into the subject, we first like to define important terms in order to contribute to a clearer understanding.

### A. Military Simulation/War Game

Within this publication the terms "military simulation" and "war game" are used synonymously and are based on a definition of Peter Perla [3] as follows:

*A military simulation/war game (also wargame) is a strategy game that deals with military operations of various types, real or fictional.*

Although there may be disagreements as to whether a particular game qualifies as a war game or not, a general consensus exists that all such games must explore and represent some feature or aspect of human behaviour directly bearing on the conduct of war, even if the game subject itself does not concern organized violent conflict or warfare [9].

## B. Taxonomy

*A taxonomy is the hierarchical structuring of a knowledge field into main groups and subcategories [4].*

According to Lindqvist and Jonsson, a taxonomy should focus on the following, ideal properties [4]:

- The categories have to be mutually exclusive; no overlapping between the categories
- Clear and unambiguous classification criteria; a repeated classification must produce the same results
- Comprehensible and useful
- Comply with established terminology

## III. TAXONOMY

Within this Section, our taxonomy and its components are described in detail (see Figure 1 to see the entire taxonomy). Following the most common sub-division, our taxonomy as well as this Section is further sub-divided into two sub-sections reflection functional and non-functional aspects [10]. Functional criteria determine *what* a product shall do. In contrast to this is the non-functional category. Here, the *properties* a product shall have, are determined.

### A. Functional Criteria

The category of functional criteria is again split into two elements: (i) General criteria for simulators and (ii) military specific functional criteria. Individual categories are as follows:

*1) General Criteria:*

*Simulation Paradigm*: This category reflects the major paradigms of simulation modeling which are [11,12]: System Dynamics, Discrete Event and Agent Based (see Table I for a short overview). Due to space limitations, no complete introduction can be done at this point. For this purpose, we like to point to [11,12] for instance.

*Time Handling* describes the 'use of time'. A simulation may have a specific duration (expressed, for example, in a maximum playing time in *hours:minutes:seconds* or number of rounds) or be of an indefinite period. Furthermore, there are simulations in which the parties operate in parallel (*simultaneously*) or sequentially (*non-simultaneous*). In the first case, especially in the military, the question whether a simulation has a (quasi) real-time capability is highly relevant.

*Geography* also plays an important role for the military. With this category, however, we do not refer to the topography (desert, city, etc.); this is done below in the category 'topography'. Here, the core issue is whether the simulation includes any geographical aspects and if so, which aspects the pitch has (raster vs. vector, etc.).

*Parametrizability:* As a method of customizing, parametrization allows to enable or disable parts of the simulator by setting parameters [13]. This allows the user to control functions and the execution of the simulator and to modify the simulator according to own ideas.

*Replay Capability* refers to the ability to record and replay of situations and events similar to a digital video recorder [14].

*History:* For the development of the concrete simulation, two different approaches can be taken: Real historical data can be used or a fictitious scenario can be adducted, if only incomplete or even no data at all is available.

*Economy* can be differentiated into dealing with economic aspects or their non-consideration (negligence).

*Game Theory* in general describes multi-person decision scenarios where each player chooses actions which result in the best possible rewards for themselfes, while anticipating the rational actions from other players [14]. Game theory comprises two branches (*game forms*): Cooperative game theory (CGT) and non-cooperative game theory (NCGT) [15-17]. CGT models how agents compete and cooperate as coalitions in unstructured interactions to create and capture value [17]. NCGT models the actions of agents, maximizing their utility in a defined procedure, relying on a detailed description of the moves and information available to each agent [16]. CGT abstracts from these details and focuses on how the value creation abilities of each coalition of agents can bear on the agents' ability to capture value. CGT can thus be called coalitional, while NCGT is procedural [17]. A second division is performed by the *number of stages*. A one-shot game in which each player chooses his plan of action and all players' decisions are made simultaneously is called "static game". This means when choosing a plan of action each player is not informed of the plan of action chosen by any other player [15]. A game with more than one stage is known as "dynamic game" [18]. It can be considered as a sequential structure of the decision making problems encountered by the players in a static game [15]. The sequences of the game can be either finite, or infinite [15]. Furthermore, a subdivision in relation to the *handling of information* is common [19]. A *perfect information game* is a game in which each player is aware of the moves of all other players that have already taken place [15]. Examples of perfect information games are: chess, tic-tac-toe, and go. A game where at least one player is not aware of the moves of at least one other player is called an *imperfect information game* [15]. A *complete information game* is a game in which every player knows both the strategies and payoffs of all players in the game, but not necessarily the actions [15]. This term is often confused with that of perfect information games but is distinct in the fact that it does not take into account the actions each player have already taken. *Incomplete information games* are those in which at least one player is unaware of the possible strategies and payoffs for at least one of the other players [15]. *Dominance* considerations can help to find solution strategies. The concept of *dominant strategy* refers to identifying a series of actions that is better than all the other options, regardless of what the other players do. The *Nash equilibrium*, describes - in non-cooperative games - a combination of strategies, where each player selects just one strategy from which it does not make sense for any player to select an alternative strategy (e.g., see [20] for more details). Finally, *actors* are also very relevant. Do the actors behave rational or irrational? Are the actors adaptable? How many actors/players are there?

TABLE I. COMPARISON OF THE MAJOR SIMULATION PARADIGMS (BASED ON [11,12])

| Simulation Paradigm | System Dynamics | Discrete Event | Agent Based |
|---|---|---|---|
| *Abstraction Layer* | High Abstraction | Low to Middle Abstraction | Used across all abstraction layers |
| *Elements/ Components* | Consists of Stock-and-Flow diagrams and Feedback loops | Main elements: Entities and Resources (Passive Objects) Flowchart Blocks (Queues, Delays, etc.) drive the model | Agents may model active objects of very diverse nature and scale Individual behavior rules Direct or indirect interaction Environment models |

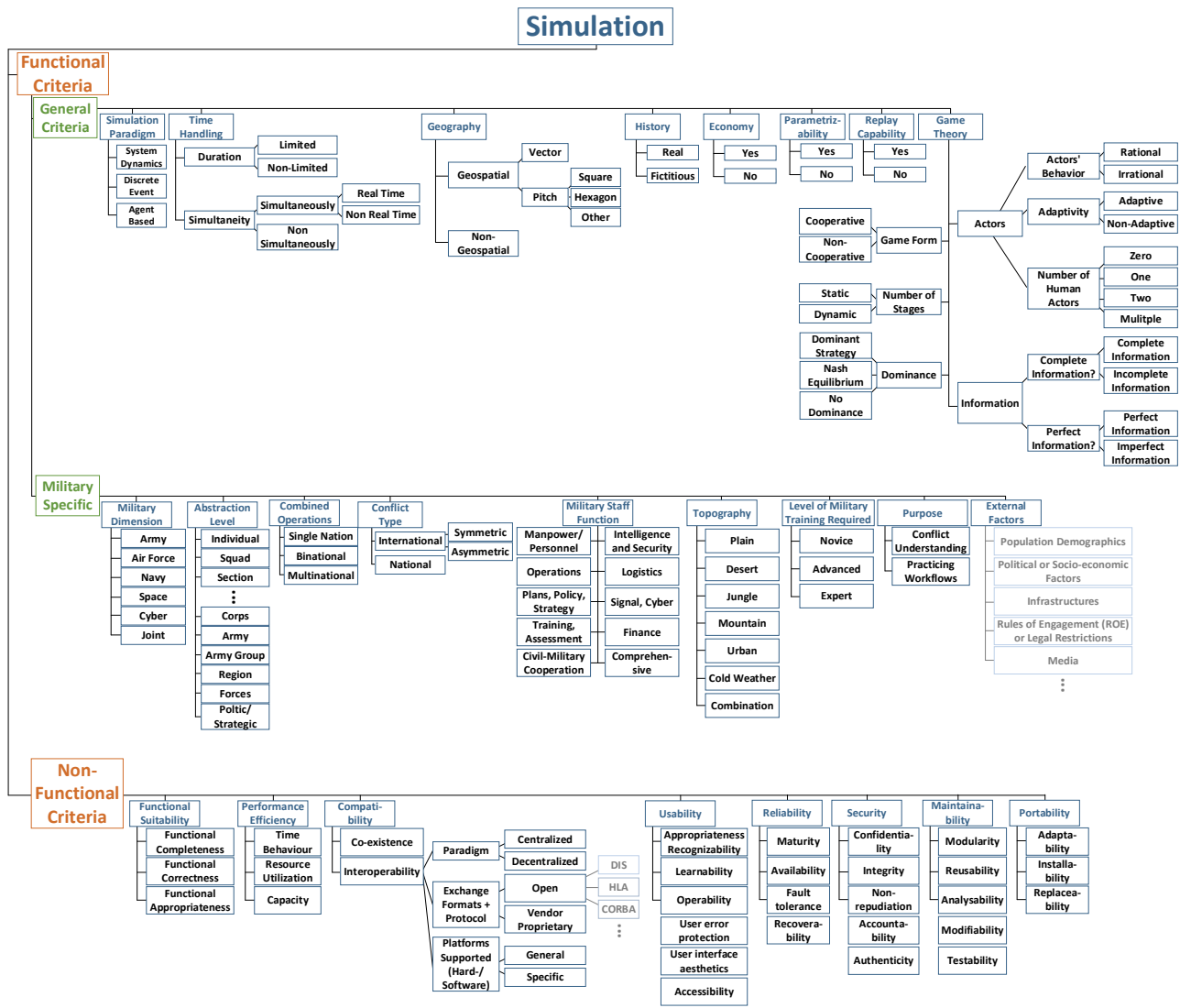

Figure 1. Taxonomy for the Assessment of Military Simulations

*2) Military Specific Criteria:*
*Military Dimension*: The division of forces into branches with specific defense material/training/capabilities/purpose etc. probably represents the best known type of division. The classic division after the first World War distinguishes in the three branches: (i) Army or land forces, (ii) Air Force or Air Defense and (iii) Navy or naval forces. Recently, these three dimensions of warfare are usually supplemented by (aero)-space and cyber space [21].

*Abstraction Level:* Military hierarchies are another distinctive feature of armed forces. In general, a differentiation is made in the following levels [22]: Individual, fireteam, squad, section, platoon, company/battery, battalion/squadron, regiment, brigade, division, corps, army group, region and the entire forces. Above this operational layer, there is the political-strategic layer. This layer can hardly be divided into further sub-components (at least without risking general acceptance).

*Combined Operation:* Military operations can be performed solely by one nation or with one resp. more partners.

*Conflict Type:* Traditionally, military conflicts are classified into two major groupings [23]: international wars and civil conflicts. International wars are defined as those in which a territorial state is engaged in a war with another state involving regular armed forces on both sides. Civil conflicts are conducted between a state and a group within its borders.

*Military Staff Function:* Military organizations are normally structured with directorates for manpower and personnel, intelligence, operations, logistics, plans, and communications systems forming the core [24]. The so-called "'continental staff system'", which forms the basis for this publication, and which is used by the overall majority of all NATO states, expands the concept by introducing three more directorates [25,26]: (i) training, (ii) finance and contracts (also known as "resource management") and (iii) Civil-Military Co-operation (CIMIC).

*Topography:* While for Air Force, Navy and especially for Aerospace, the topography is usually of subordinate relevance, topography is of great significance for the Army. Consequently terrain analysis plays an important role for the Army (see for example [27,28]). Depending on the type of terrain, several guidelines have been developed and need to be taken into account. For this purpose, for example, the US military has developed regulations (so-called Field Manuals) for (i) desert operations [29], (ii) jungle operations [30], (iii) mountain operations [31], (iv) Military Operations on Urbanized Terrain [32] and (v) Cold Weather Operations [33]. This division is also found in our taxonomy.

*Level of Military Training Required:* This categories reflects how much knowledge/military training a user must have before it makes sense that he/she uses the simulator. This, however, does neither imply knowledge about simulation in general/other simulators nor knowledge about IT.

*Purpose:* This aspect deals with the question whether the main sense of a simulator consists in the idea to facilitate conflict understanding (with all facets), or whether the training of military workflows (such as the evaluation of the current situation and the military planning process) is the primary focus .

*External Factors:* Battlefield's Effects are also dependent on a couple of external factors which have no direct military origin. This includes (but is not limited to) politics, civilian press, local population, demographics and other socioeconomic factors, infrastructures, legal restrictions such as rules of engagement (ROE) or media [28].

### B. Non-funtional Criteria

Whereas current practices primarily focuses on functional requirements, considerations other than the function (e.g., safety; security; maintainability) are usually relegated into a category called "non-functional requirements" [34]. ISO/IEC /IEEE 24765 defines this term as "*a software requirement that describes not what the software will do but how the software will do it*" [35]. In the field of software engineering in particular, the ISO 25000 is used to ensure that quality [33]. This is also due to the fact that the ISO/IEC 25000 family of standards suggests an objective evaluation with the help of a quality model, through which an abstract quality attribute is decomposed into measurable characteristics [36,34].

For this the ISO 25010 standard contains eight categories, which we are re-using here [37]:

*Functional Suitability* refers to the degree to which a product or system provides functions that meet stated and implied needs when used under specified conditions; it is composed of the following sub-characteristics:
- *Functional Completeness:* The degree to which the set of functions covers all the specified tasks and user objectives. In other words, if the military simulations contains all the required functions.
- *Functional Correctness:* The degree to which the military simulation provides the correct results with the needed degree of precision (as defined in the conceptual model).
- *Functional Appropriateness:* The degree to which the functions (within the simulation) facilitate the accomplishment of specified tasks and objectives.

*Performance Efficiency* represents the performance of the simulation relative to the amount of resources used. The following sub-characteristics specify this in more detail:
- *Time Behaviour:* The degree to which the response and processing times and throughput rates of the simulation meet the requirements.
- *Resource Utilization:* The degree to which the amounts and types of resources used by the simulation meet the requirements.
- *Capacity:* The degree to which the maximum limits meet the requirements.

*Compatibility* describes the degree to which the simulation can exchange information with other products, systems or components, and/or perform its required functions, while sharing the same hardware or software environment. This characteristic is composed of the following sub-characteristics:
- *Co-existence:* The degree to which the simulation can perform its required functions efficiently while sharing a common environment and resources with other products, without detrimental impact on any other product.

- *Interoperability:* The degree to which two or more simulations can exchange information and use the information that has been exchanged. Over the recent years, in the area of military simulations, there is an ongoing trend observable that simulations have to have the ability to be linked together [38]. Inspired by the concept "train as you fight", future multinational contingents should start to jointly operate at a very early stage of training. For these reasons, the authors have carried out a refinement of this category. Interoperability paradigm here refers to the question whether a distributed simulation has a central control or not (centralized vs. decentralized). Furthermore, exchange formats and protocols play an important role. Most suitable are publicly available standards, like Distributed Interactive Simulation (DIS) [39] or High-Level Architecture (HLA) [40]. Vendor proprietary approaches however, are clearly less useful in terms of interoperability.

*Usability* corresponds to the degree to which a military simulator can be used to achieve specified goals w.r.t. effectiveness, efficiency and satisfaction in a specified context of use. This characteristic is composed of the following sub-characteristics:

- *Appropriateness Recognizability:* The degree to which users can recognize whether a military simulator is appropriate for their needs.
- *Learnability:* The degree to which a military simulator can be used to achieve specified goals of learning with effectiveness, efficiency, freedom from risk and satisfaction in a specified context of use.
- *Operability:* The degree to which a military simulator has attributes that make it easy to operate and control.
- *User error protection:* The degree to which a military simulator protects users against making errors.
- *User interface aesthetics:* The degree to which the user interface of a military simulator enables pleasing and satisfying interaction for the user.
- *Accessibility:* The degree to which a military simulator can be used by people with the widest range of characteristics and capabilities to achieve a specified goal in a specified context of use.

*Reliability* addresses the question whether a military simulator performs specified functions under specified conditions for a specified period of time. This characteristic is composed of the following sub-characteristics:

- *Maturity:* The degree to which a military simulator meets needs for reliability under normal operation.
- *Availability:* The degree to which a military simulator is operational and accessible when required for use.
- *Fault Tolerance:* The degree to which a military simulator operates as intended despite the presence of hardware or software faults.
- *Recoverability:* The degree to which, in the event of an interruption or a failure, a military simulator can recover the data directly affected and re-establishes the desired state of the system.

*Security* describes whether a military simulator protects information. This characteristic is composed of the following sub-characteristics:

- *Confidentiality:* The degree to which a military simulator ensures that data is accessible only for those authorized to have access to.
- *Integrity:* The degree to which a military simulator prevents unauthorized access to, or modification of, computer programs or data.
- *Non-repudiation:* The degree to which actions or events can be proven to have taken place, so that the events or actions cannot be repudiated later.
- *Accountability:* The degree to which the actions of an entity can be traced uniquely to the entity.
- *Authenticity:* The degree to which the identity of a subject or resource can be proved to be the one claimed.

*Maintainability* represents effectiveness and efficiency with which a military simulator can be modified to improve it, correct it or adapt it to changes in environment, and in requirements. This characteristic is composed of the following sub-characteristics:

- *Modularity:* The degree to which a military simulator is composed of discrete components such that a change to one component has minimal impact on other components.
- *Reusability:* The degree to which an asset can be used in more than one system, or in building other assets.
- *Analyzability:* The degree of effectiveness and efficiency with which it is possible to assess the impact on a military simulator of an intended change to one or more of its parts, or to diagnose a product for deficiencies or causes of failures, or to identify parts to be modified.
- *Modifiability:* The degree to which a military simulator can be effectively and efficiently modified without introducing defects or degrading existing product quality.
- *Testability:* The degree of effectiveness and efficiency with which test criteria can be established for a military simulator and tests can be performed to determine whether those criteria have been met.

*Portability* refers to effectiveness and efficiency with which a military simulator can be transferred from one hardware, software or other operational or usage environment to another. This characteristic is composed of the following sub-characteristics:

- *Adaptability:* The degree to which a military simulator can effectively and efficiently be adapted for different or evolving hardware, software or other operational or usage environments.
- *Installability:* The degree of effectiveness and efficiency with which a military simulator can be

successfully installed and/or uninstalled in a specified environment.
- *Replaceability:* The degree to which a product can replace another specified software product for the same purpose in the same environment.

IV. CASE STUDY: NETLOGO REBELLION

The purpose of this section is to briefly introduce Net-Logo Rebellion and then to classify it using the taxonomy - due to the page limit, unfortunately, only partly. The key aspect within Rebellion, which is running on top of the platform NetLogo (a multi-agent programmable modeling environment [41]), is the question when and if a subjugated population rebels against a central authority [42]. To this end, the population wanders around randomly. If their level of grievance against the central authority is high enough, and their perception of the risks involved is low enough, they openly rebel [42]. A separate population of police officers ("cops"), acting on behalf of the central authority, seeks to suppress the rebellion. To this end, the cops also wander around randomly and arrest people who are actively rebelling [42].

The conceptual model on which the simulation has been created is described in Figure 2. The behavior of the agents (civil population/cops) can be influenced by variables, which can be set by the user. These variables are:

- Proportion of police in contrast to the total population
- Proportion of civilian in the total population
- Sight of police (visibility)
- Legitimacy of the central authority
- Maximum time in prison
- The question whether civilians and police officers are able to move

Additional variables are either set randomly for the entire simulation (for example the system constant k, which serves to avoid specific side effects in the simulation; see [42] for more details) or set randomly for each individual civilians (here: Risk Aversion resp. Trust in Government).

Focusing on functional criteria (due to space limitations), Rebellion can be classified as follows. First, the simulation paradigm of NetLogo is classified as agent based (interactions of autonomous agents). The time handling of Rebellion is of a limited duration, simultaneously in non real-time. The geography is non-geospatial, the history fictitious. Economical aspects are not taken into consideration within Rebellion, therefore economy is not applicable. The simulation is parametrizable as well as replayable. However, when replayed, different solutions will occur due to the use of random variables. Wrt. game theory, no dominance can be applied. Next to this, every player knows both the strategies and payoffs of all players in the game *complete information game*. Depending on the visibility (set by the user), each player is aware of the moves of all other players that have already taken place (*perfect information game*) or not (*imperfect information game*). Actors are not adapting and behave according to Figure 2, which the authors are judging as irrational, since a rational behavior of civilians would for instance take into account how many cops are in their "local environment". The model itself has a punctuated equilibrium (theory in evolutionary biology) which should not be confused with the Nash equilibrium (see [42] for more details).

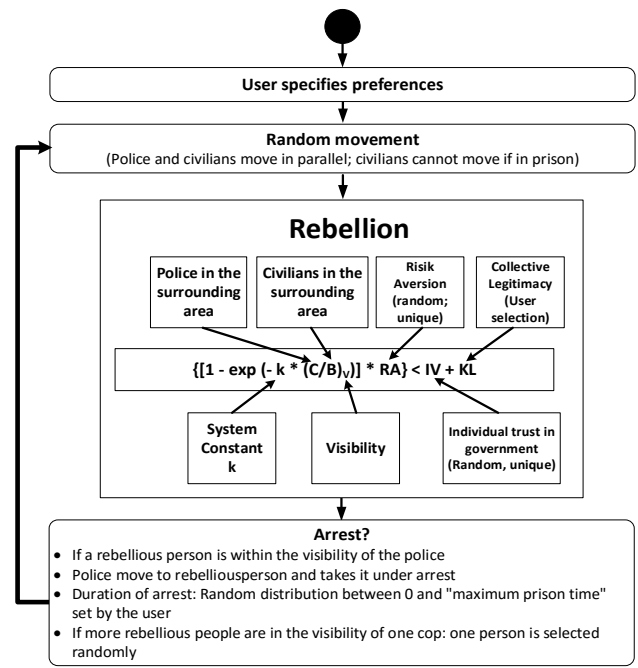

Figure 2. Conceptual Model of Rebellion

Referring to military specific criteria, Rebellion can be applied to asymmetric, international conflicts on an individual abstraction level. Wrt. the military branch, NetLogo Rebellion focusses on army operations of one single nation. The military staff function of "operations" with a plain topography is used. Because of the user-friendly interface with limited parameters as well as the simple simulation environment, only a novice level of military training is required. The purpose of Rebellion is "practicing workflows", as its application can be used to optimize patrols in monitored areas.

V. CONCLUSION

Simulations are an important aspect for the military in general and the military planning process as well as the estimation of endangerments in specific. Therefore, numerous different solutions and techniques have been developed over the past years. However, the selection of a specific simulation systems (for a use case) as well as the further development of systems can be hampered because of a missing general classification. To overcome this shortcoming, this publication presented a first contribution to the development of a commonly applicable taxonomy for military simulations. To demonstrate the use of our new taxonomy, its application to NetLogo Rebellion was presented. At this point we like to recall the fundamental purpose again. The aim of the taxonomy is to allow an assessment of military simulators using different categories, without the need for the assessor to be a qualified expert in the particular category. Based upon this consideration, individual categories may sometimes appear to be in-comprehensive (at least in the eyes of an expert). This applies for example applies to the category of "game theory", where individual categories are likely to be divided into more sub-categories (if experts are involved). Furthermore, our sub-division is based on "majority opinions". The latter, however, already implies that often a possible categorization can be made in different ways (e.g., concerning the question of what is meant by a "conflict").

Based on the presented work, we want to motivate the discussion of our taxonomy, enabling an improved selection, application and development of military simulation systems.

ACKNOWLEDGMENT

At this point, the authors want to thank Prof. Beckmann, Prof. Meissner and Prof. Bayer from the Helmut-Schmidt-Universität.

This work was partly funded by FLAMINGO, a Network of Excellence project (ICT-318488) supported by the European Commission under its Seventh Frame-work Programme.